\documentclass[11pt,twoside]{article}
\usepackage{./asp2014}

\aspSuppressVolSlug
\resetcounters

\begin{document}

\title{Correlated X-ray and optical variability in intermediate polars during their outbursts}
\author{V.~V.~Neustroev,$^{1,2}$ S.~Tsygankov,$^3$ V.~Suleimanov,$^4$ and G.~Sjoberg$^{5,6}$
\affil{$^1$Finnish Centre for Astronomy with ESO (FINCA), University of Turku, Finland; \email{vitaly@neustroev.net}}
\affil{$^2$Astronomy research unit, University of Oulu, Finland}
\affil{$^3$Tuorla Observatory, University of Turku, Finland}
\affil{$^4$Institut f\"ur Astronomie und Astrophysik, Universit\"at T\"ubingen, Germany}
\affil{$^5$The George-Elma Observatory, Mayhill, New Mexico, USA}
\affil{$^6$American Association of Variable Star Observers, Cambridge, USA}}

\paperauthor{Neustroev, V. V.}{vitaly@neustroev.net}{0000-0003-4772-595X}{University of Turku}{Finnish Centre
for Astronomy with ESO (FINCA)}{Turku}{}{FIN-21500}{Finland}
\paperauthor{Sample~Author2}{Author2Email@email.edu}{ORCID_Or_Blank}{Author2 Institution}{Author2 Department}{City}{State/Province}{Postal Code}{Country}
\paperauthor{Sample~Author3}{Author3Email@email.edu}{ORCID_Or_Blank}{Author3 Institution}{Author3 Department}{City}{State/Province}{Postal Code}{Country}

\begin{abstract}
We present a study of the evolution of the optical and X-ray fluxes during outbursts of two short-period
cataclysmic variables, the confirmed intermediate polar CC~Scl and the intermediate
polar candidate FS~Aur. We found that the X-ray and optical light curves are well correlated in both
objects, although the amplitudes of outbursts in X-rays are smaller than those in the optical. The ratio
of the outburst amplitudes in X-rays and the optical in both objects is close to $\sim$0.6.
This is significantly higher than was observed during the outburst of the non-magnetic dwarf nova U~Gem,
in which this ratio was only $\sim$0.03. The obtained data also suggest that the dependence between
the X-ray and optical fluxes must steepen significantly toward very low accretion rates and very low fluxes.
Similarities in the behaviour of CC~Scl and FS~Aur indicate strongly the magnetic nature of the white
dwarf in FS~Aur.
\end{abstract}

\section{Introduction}

Cataclysmic Variables (CVs -- for a comprehensive review, see \citealt{Warner}) are close interacting
binary systems in which a white dwarf (WD) accretes
material from its late-type low mass companion. In the absence of a strong magnetic field, the material
transferred from the donor star forms an accretion disc around the WD and gradually spirals down onto
its surface. If the mass-transfer rate is low, the accretion disc can suffer a thermal instability
caused by hydrogen ionisation, resulting in outbursts. Dwarf novae (DN) are
an important subset of CVs, which undergo such outbursts. Intermediate polars (IP) are another subset
of CVs in which the magnetic field of the WD is strong enough to disrupt the inner accretion disc and
force the accreting material to flow along field lines on to one or both magnetic poles. Outbursts can
also arise in the truncated discs of IPs, although only a few such objects are known (e.g., YY~Dra,
CC~Scl, GK~Per).

The difference in the strength of the magnetic field of the WD in non-magnetic DNe and IPs determines the
time evolution of the X-ray radiation during their outbursts. The X-ray emission of non-magnetic CVs is
believed to originate from regions very close to the WD surface, most likely in a boundary layer. During
an outburst, the boundary layer transitions from optically thin to optically thick, which may lead to a
suppression of the X-ray flux. Indeed, only a very few non-magnetic DNe are known in which the X-ray flux
increases during an outburst (e.g., U~Gem -- \citealt{Mattei}). In IPs the accretion occurs onto the magnetic
poles of the WD, and the boundary layer does not exist. Thus the X-ray flux during an IP outburst is expected
to increase. Indeed, such a behaviour was observed in the above-mentioned outbursting IPs \citep{Szkody,Woudt,GK_Per}.
Here we present a comparison of the evolution of the optical and X-ray fluxes during outbursts of two CVs
with similar orbital periods, the confirmed IP CC~Scl and the IP candidate FS~Aur.


\section{Background and the data}

\subsection{CC Sculptoris}

CC~Scl was discovered as the ROSAT source RX J2315.5$-$3049 \citep{Voges} and later classified as
a dwarf nova \citep{Schwope}, although no outbursts have yet been detected. \citet{Ishioka} observed
two outbursts of CC~Scl, one of which appeared to be a superoutburst due to the detection of superhumps,
thus classifying CC~Scl as a SU~UMa-type dwarf nova. From time-resolved spectroscopy, \citet{Chen}
derived an orbital period of 84.1~min. In 2011, \citet{Woudt} performed optical spectroscopic, high-speed
photometric and {\it Swift} X-ray observations of the superoutburst of the object. They detected WD
spin modulations with a period of 6.49~min which are seen in both optical and X-ray light curves
only during the outburst.  Thus, CC~Scl is a confirmed intermediate polar with one of the shortest
orbital periods among IPs.

Figure 11 in \citet{Woudt} shows that an X-ray count-rate from CC~Scl increased in the outburst.
In order to compare the changes in the optical and X-ray energy ranges, we used the observations published
in \citet{Woudt}. We note that the X-ray spectrum of CC~Scl has changed significantly during the outburst
and the following decline, mostly because of changes in the absorption column \citep{Woudt}. For this
reason, in our work we used unabsorbed X-ray fluxes in the energy band 0.3--10 keV. They were calculated
using the partially-covered cooling flow model described in \citet{Woudt}. These fluxes were not published
in the paper, but were provided to us by K.~L.~Page, whose help is very much appreciated. We also used
the $V$ observations of F.-J.~Hambsch which were published in \citet{Woudt} and stored in the AAVSO database.
In addition, we extracted the 1-d averaged SAAO observations from figure 1 in their paper. The $V$ magnitudes
were then converted into fluxes and interpolated to the times of the X-ray observations.

\subsection{FS Aurigae}

FS Aur is a peculiar CV showing multiple periodic photometric and spectroscopic variabilities \citep{T03,T07}.
Its orbital period of 85.7 min is determined from the radial velocity variations of emission lines \citep{Thorstensen,N02},
and is close to that of CC~Scl. The puzzling behaviour of FS~Aur is explained within the frame of an enhanced
IP scenario with a precessing, fast-rotating magnetically accreting WD \citep{T07}. The spin period of the WD
is expected to be of the order of 50--100~s. This hypothesis has received an observation confirmation after
finding strong indications that many observed properties of FS Aur closely resemble those of other IPs \citep{N13}.
However, the coherent spin modulations are not yet discovered, thus the IP status of FS~Aur is still unconfirmed.

FS~Aur shows regular outbursts with a total duration of about 4~d and of relatively low amplitude of $\sim$2 mag,
with recurrence time of 18.1$\pm$2.5~d \citep{N12}. Most of the observed outbursts had very similar light curves. It
is interesting that FS~Aur also exhibits strong variability even in quiescence. For example, \citet{N12} reported
that during a 5-month observational campaign in 2010-2011 the average quiescent level of the system has increased
by 0.3--0.4 mag. Sometimes FS~Aur experiences drops in brightness by about 1.5 mag \citep{T07}. \citet{Chavez}
showed that the optical brightness of FS~Aur in quiescence varies within a wide range (17.4--15.2~mag) and these
modulations of the light curve are possibly periodic with a $\sim$900-d period.

FS~Aur is a rather hard and relatively bright X-ray source. \citet{N13} reported the X-ray observations of FS~Aur
obtained with {\it ROSAT}, {\it Chandra} and {\it Swift} in quiescence at different optical brightness level. They
noted that the X-ray flux varies proportionally to the optical flux. In order to study this effect in more detail,
in this work we additionally use several new observations obtained recently with {\it Suzaku} \citep{N14} and {\it Swift}.
Thus, our quiescence data cover about a 1-mag range of optical magnitudes.

In January 2015, we were lucky to catch FS~Aur at the onset of a new outburst, the development of which we traced
in detail from the maximum to quiescence. We obtained 11 nights of optical photometry using the 0.35~m Celestron
C14 robotic telescope, located at New Mexico Skies in Mayhill, New Mexico. The data were taken with an SBIG ST-10XME
CCD camera and the standard Johnson $V$ filter. We were able to start our ToO {\it Swift} X-ray observations at the
optical light maximum and we observed until the system returned back to quiescence. We note that an X-ray count-rate
had also increased in the outburst, similarly to CC~Scl. To calculate the X-ray fluxes in the energy band 0.3--10 keV,
we used the three-temperature cooling flow model described in \citet{N13}.

\articlefiguretwo{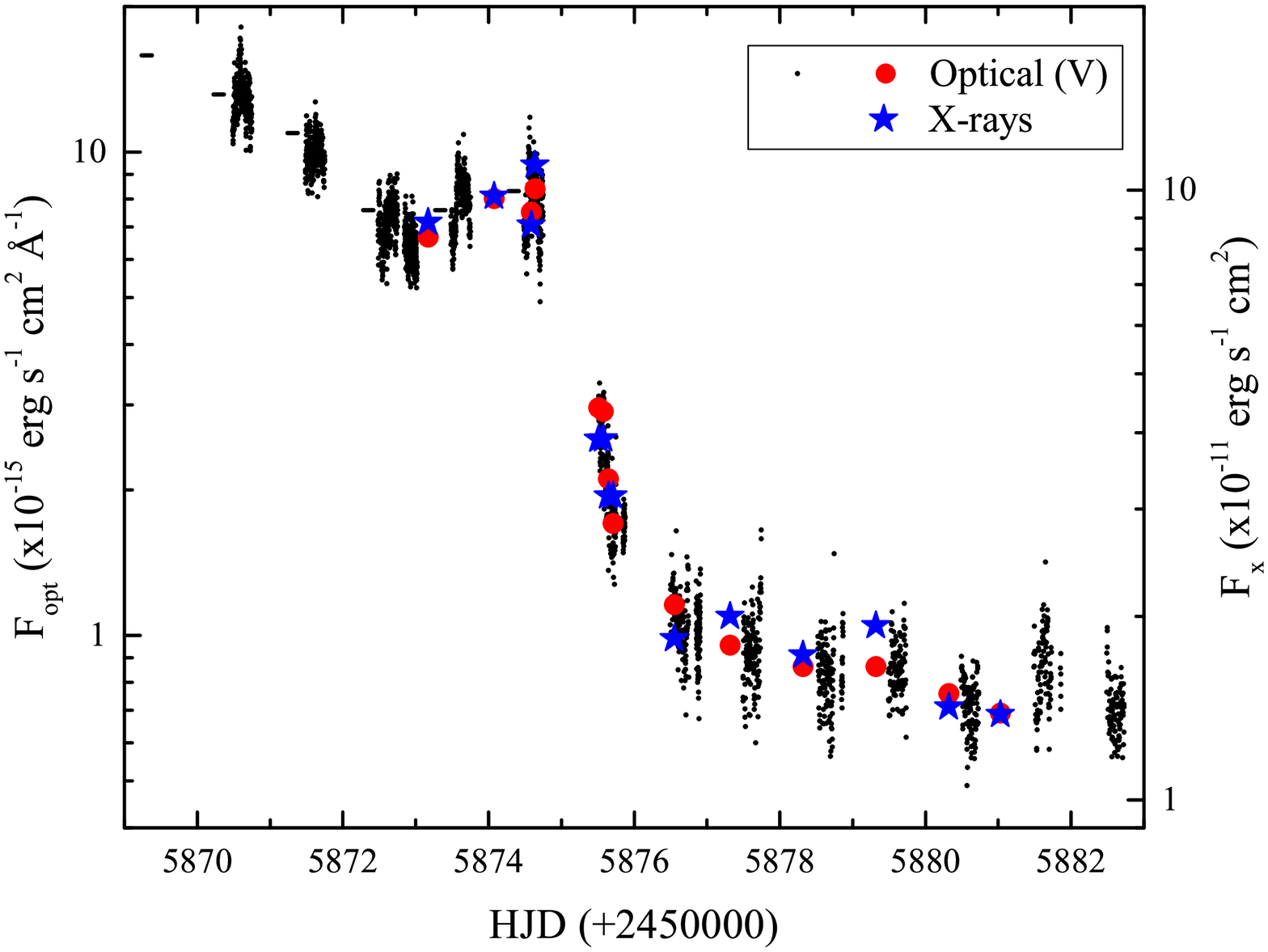}{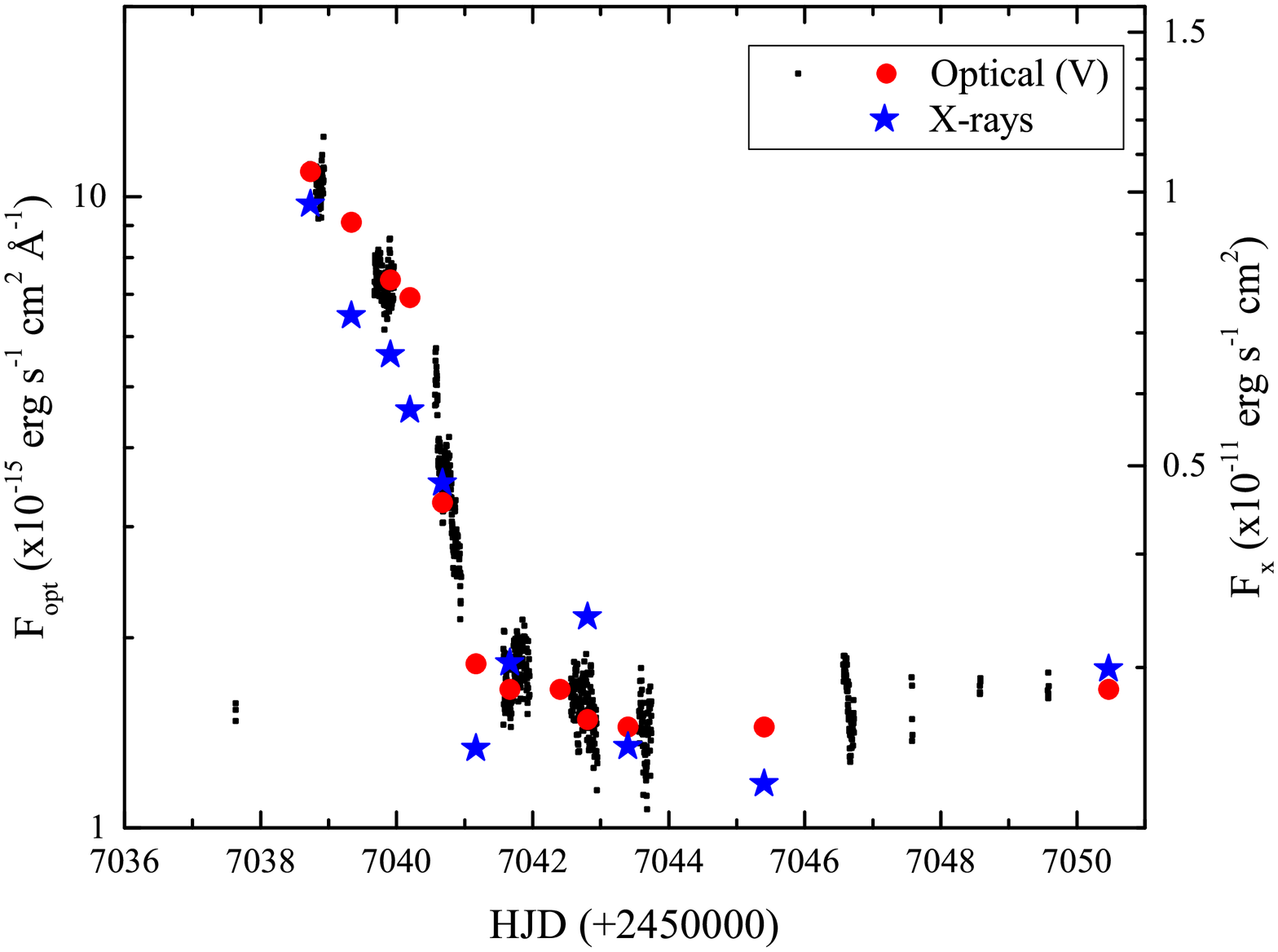}{Fig:LCs}{The optical and X-ray light curves
of the outbursts of CC~Scl (left) and FS~Aur (right).}

\articlefigure{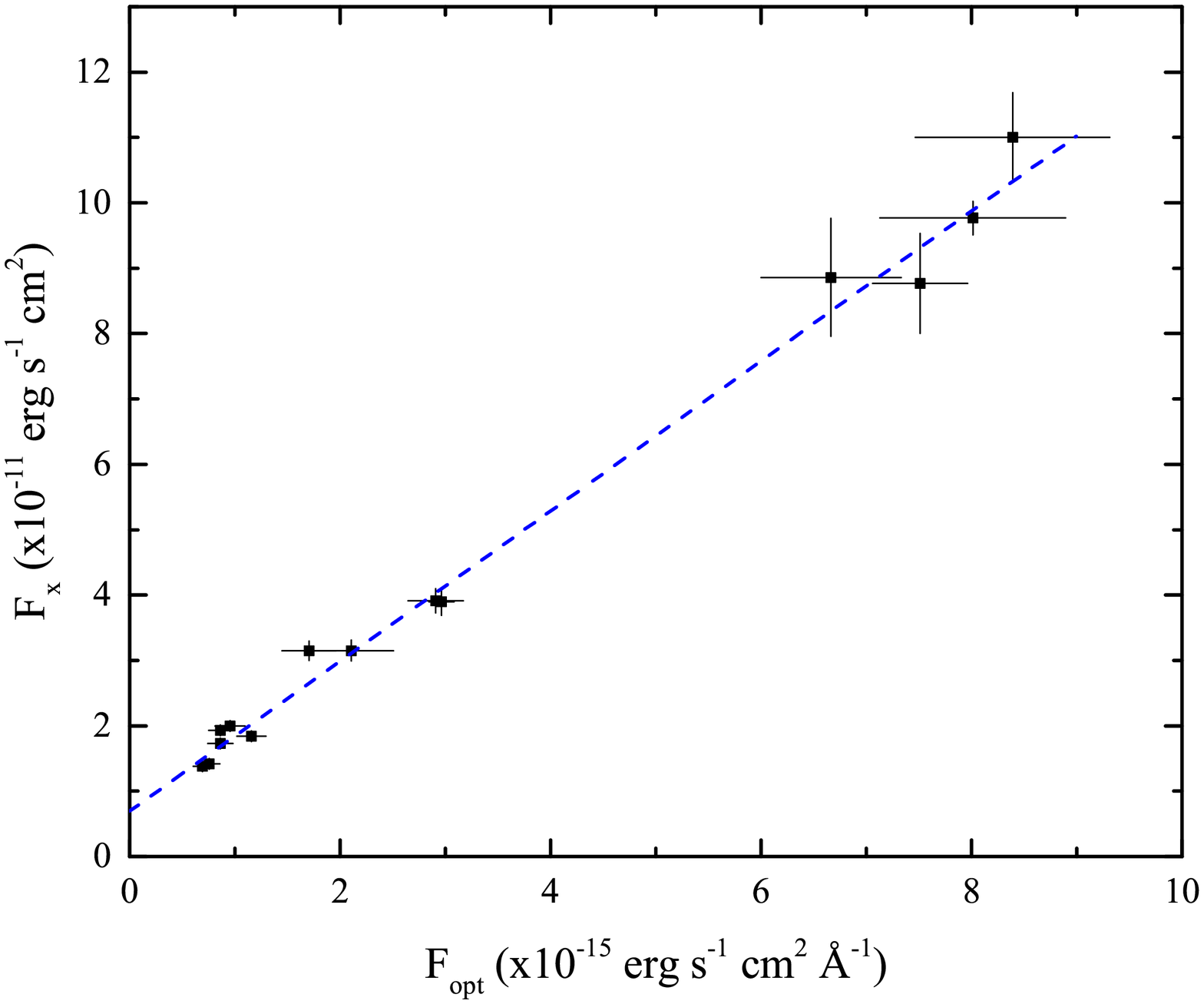}{Fig:CC_XOpt}{The optical/X-ray flux diagram for CC~Scl.}
\articlefigure{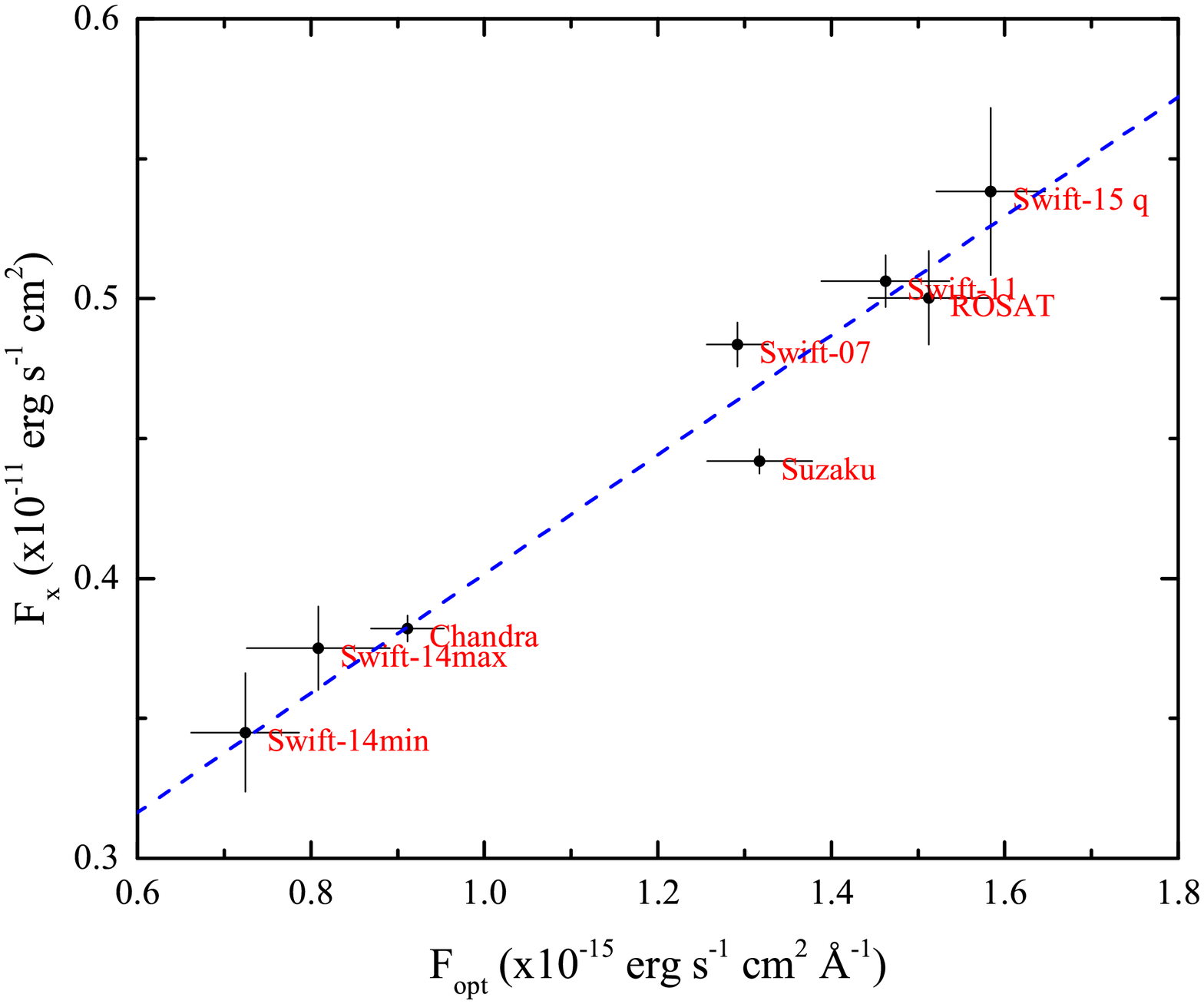}{Fig:FS_XOpt1}{The optical/X-ray flux diagram for FS~Aur in quiescence.}
\articlefigure{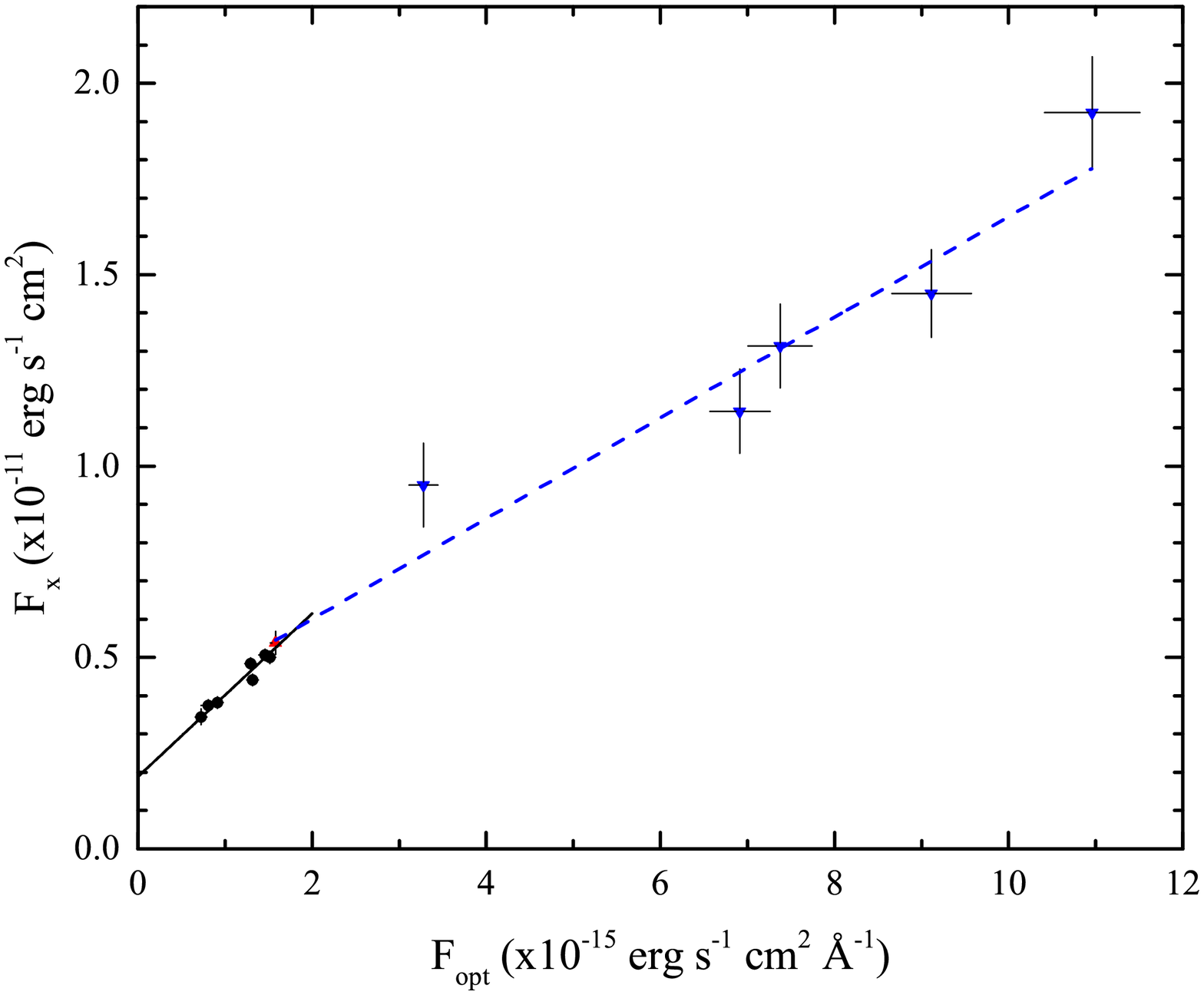}{Fig:FS_XOpt2}{The optical/X-ray flux diagram for FS~Aur in quiescence
(black diamonds) and in the outburst (blue triangles).}

\section{Results}

Figure~\ref{Fig:LCs} shows the optical and X-ray light curves of the outbursts of CC~Scl (left-hand panel)
and FS~Aur (right-hand panel), and in Figs.~\ref{Fig:CC_XOpt}, \ref{Fig:FS_XOpt1} and \ref{Fig:FS_XOpt2}
we show the optical/X-ray flux diagrams. The morphology of the optical and X-ray light curves of CC~Scl
is very similar, resulting in a nearly linear dependence between the X-ray and optical fluxes
(Fig.~\ref{Fig:CC_XOpt}). However, the outburst amplitude in the X-rays is less than in the optical.
In order to characterize the difference in amplitudes of flux variations, we introduce a parameter $R$,
the rate of changes of the X-ray and optical fluxes:
\begin{equation*}
R=\frac{F_{x,max}}{F_{x,min}} / \frac{F_{opt,max}}{F_{opt,min}} \,
\end{equation*}
where $F$ are the maximal and minimal X-ray and optical fluxes. For CC~Scl, $R$$\approx$0.62.
An interesting feature of the optical/X-ray flux diagram is that the fitted line crosses the zero optical
flux axis with a non-zero X-ray flux. This indicates that the dependence between the X-ray and optical
fluxes must steepen significantly toward very low accretion rates (and very low fluxes).

The optical and X-ray fluxes of FS~Aur in quiescence show a similar dependence with $R_{qui}$$\approx$0.71
(Fig.~\ref{Fig:FS_XOpt1}). The fitted line also crosses the zero optical flux axis with a non-zero, positive
X-ray flux. During the outburst of FS~Aur, the dependence between the X-ray and optical fluxes became less
steep, though not very significantly (Fig.~\ref{Fig:FS_XOpt2}). The ratio of the outburst amplitudes in
X-rays and the optical is $R_{outb}$$\approx$0.47.

Thus, we obtained that during the outbursts of the confirmed IP CC~Scl and the IP candidate FS~Aur, their
X-ray fluxes increased with a similar rate $R$. This rate is significantly higher than was observed in
those normal dwarf novae with a non-magnetic WD, which showed an increase of X-ray flux in outburst.
For example, during an outburst of U~Gem \citep{Mattei} $R$ was only about 0.03.


\section{Summary and Conclusion}

We investigated the relation between the optical and X-ray fluxes during outbursts of two CVs with similar
orbital periods. We found the following:
\begin{enumerate}
  \item The ratio of the outburst amplitudes in X-rays and the optical in both objects is close to $\sim$0.6,
  that is significantly larger than in the non-magnetic dwarf nova U~Gem.
  \item The dependence between the X-ray and optical fluxes must steepen significantly toward very low accretion
  rates and very low fluxes.
  \item Similarities in the behaviour of CC~Scl and FS~Aur indicate strongly the magnetic nature of
  the WD in FS~Aur.
\end{enumerate}

\acknowledgements VS thanks Deutsche Forschungsgemeinschaft (DFG)  for  financial  support  (grant  WE  1312/48-1).
This research has made use of data obtained from the {\it Suzaku} satellite, a collaborative
mission between the space agencies of Japan (JAXA) and the USA (NASA). Our research was based on X-ray observations
by NASA missions {\it Chandra} and {\it Swift} which we acknowledge. We thank Neil Gehrels for approving the Target
of Opportunity observation with {\it Swift} and the {\it Swift} team for executing the observation. We acknowledge
with thanks the variable star observations from the AAVSO International Database contributed by observers worldwide
and used in this research. We acknowledge the data kindly provided to us by our colleague K.~L.~Page.



\end{document}